\documentclass[a4paper,USenglish,cleveref,autoref]{lipics-v2019}

\usepackage{proof}
\usepackage{alltt}
\usepackage{verbatim}
\usepackage{wasysym}
\nolinenumbers

\title{Deep Static Modeling of \invokedynamic{}} 

\author{George Fourtounis}{University of Athens, Department of Informatics and Telecommunications, Greece}{gfour@di.uoa.gr}{}{}
\author{Yannis Smaragdakis}{University of Athens, Department of Informatics and Telecommunications, Greece}{smaragd@di.uoa.gr}{}{}

\authorrunning{G. Fourtounis and Y. Smaragdakis}

\Copyright{George Fourtounis and Yannis Smaragdakis}



\funding{We gratefully acknowledge funding by the European Research Council, grants 307334 (SPADE) and 790340 (PARSE), a Facebook Research and
Academic Relations award, and an Oracle Labs
collaborative research grant.}

\hideLIPIcs  

\EventEditors{Alastair F. Donaldson}
\EventNoEds{1}
\EventLongTitle{33rd European Conference on Object-Oriented Programming (ECOOP 2019)}
\EventShortTitle{ECOOP 2019}
\EventAcronym{ECOOP}
\EventYear{2019}
\EventDate{July 15--19, 2019}
\EventLocation{London, United Kingdom}
\EventLogo{}
\SeriesVolume{134}
\ArticleNo{27}

\newcommand{\ruleAnd}{\hspace{1.7em}}

\newcommand{\invokedynamic}{\texttt{invokedynamic}}
\newcommand{\invokestatic}{\texttt{invokestatic}}

\newcommand{\LambdaMetafactoryInvoke}[3]{\ensuremath{\mathit{MetafactoryInvo}({#1}, {#2}, {#3})}}
\newcommand{\LambdaMetas}{\ensuremath{L^{\lambda}}}
\newcommand{\InvokedynamicBoot}{\ensuremath{\mathit{Boot}}}
\newcommand{\intf}{\ensuremath{t^i}}
\newcommand{\VPT}[2]{\ensuremath{{#1} \mapsto {#2}}}
\newcommand{\IFPT}[2]{\ensuremath{{#1} \mapsto {#2}}}
\newcommand{\APT}[2]{\ensuremath{{#1} [*] \mapsto {#2}}}

\newcommand{\val}{\ensuremath{\mathit{val}}}
\newcommand{\Dyn}{\ensuremath{\mathit{Dyn}}}
\newcommand{\Arity}[1]{\ensuremath{\#{#1}}}

\newcommand{\AssignRet}[1]{\ensuremath{\mathit{Ret}(#1)}}
\newcommand{\ReifiedClass}[1]{\ensuremath{\mathit{Reified}_C(#1)}}
\newcommand{\ReifiedMethod}[1]{\ensuremath{\mathit{Reified}_M(#1)}}

\newcommand{\CGEB}[2]{\ensuremath{{#1} \dashrightarrow_b {#2}}}
\newcommand{\InvokedynamicCallSite}[3]{\ensuremath{\mathit{CSite}({#1}, {#2}, {#3})}}
\newcommand{\CallSiteContents}[3]{\ensuremath{\mathit{CSite}_C({#1}, {#2}, {#3})}}
\newcommand{\Target}[1]{{#1}\texttt{.target}}
\newcommand{\Frm}[2]{\ensuremath{F^{#1}_{#2}}}
\newcommand{\Act}[2]{\ensuremath{A^{#1}_{#2}}}
\newcommand{\Ret}[3]{\ensuremath{R^{#1}_{#2} = {#3}}}
\newcommand{\LambdaCaptured}[3]{\ensuremath{\mathit{Capture}({#1}, {#2}, {#3})}}
\newcommand{\ThisVar}[1]{\ensuremath{m/\texttt{this}}}
\newcommand{\IsConstructor}[1]{\ensuremath{\mathit{Constr(#1)}}}
\newcommand{\IsStatic}[1]{\ensuremath{\mathit{Static(#1)}}}
\newcommand{\mockC}[2]{\ensuremath{\mathit{mock}_c({#1}, {#2})}}
\newcommand{\mockH}[2]{\ensuremath{\mathit{mock}_h({#1}, {#2})}}
\newcommand{\TypeCompat}[2]{\ensuremath{\mathit{{#1} \leftrightarrows {#2}}}}
\newcommand{\DMHLookup}[3]{\ensuremath{\mathit{DMHLookup}({#1}, {#2}, {#3})}}
\newcommand{\ParamsReceiverShiftRight}[4]{\ensuremath{\mathit{Shift}({#1}, {#2}, {#3}, {#4})}}
\newcommand{\AsType}[2]{\ensuremath{\mathit{AsType}({#1}, {#2})}}
\newcommand{\In}{IN}
\newcommand{\Out}{OUT}
\newcommand{\Internal}{INTER}
\newcommand{\Rule}[1]{\textsc{#1}}
\newcommand{\LambdaObject}[4]{\ensuremath{\mathit{Lambda}({#1}, {#2}, {#3}, {#4})}}
\newcommand{\RuleMHMETHOD}{\Rule{MHMethod}}
\newcommand{\RuleMHCGE}{\Rule{MHCGE}}
\newcommand{\RuleMHARGS}{\Rule{MHArgs}}
\newcommand{\RuleRETH}{\Rule{RetH}}
\newcommand{\RuleMHCONSTR}{\Rule{MHConstr}}
\newcommand{\RuleASTYPE}{\Rule{AsType}}
\newcommand{\RuleMHLOOKUP}{\Rule{MHLookup}}
\newcommand{\RuleMHLOOKUPC}{\Rule{MHLookupC}}
\newcommand{\RuleUNREFLECT}{\Rule{Unreflect}}
\newcommand{\RuleFIND}{\Rule{Find}}
\newcommand{\RuleMTYPE}{\Rule{MType}}
\newcommand{\RuleBARGS}{\Rule{Bargs}}
\newcommand{\RuleBARGSZ}{\Rule{Bargs0}}
\newcommand{\RuleBARGSV}{\Rule{BargsV}}
\newcommand{\RuleRETB}{\Rule{RetB}}
\newcommand{\RuleCSITE}{\Rule{CSite}}
\newcommand{\RuleCALLSITECONTENTS}[1]{\Rule{CSite{#1}}}
\newcommand{\RuleMHCGEI}{\ensuremath{\Rule{MHCGE}_\textsc{Dyn}}}
\newcommand{\RuleMETAFACTORY}{\Rule{Metafactory}}
\newcommand{\RuleLAMBDA}{\Rule{Lambda}}
\newcommand{\RuleCAPTURE}{\Rule{Capture}}
\newcommand{\RuleLARGS}{\Rule{LArgs}}
\newcommand{\RuleCAPTARGS}{\Rule{CaptArgs}}
\newcommand{\RuleLAMBDATHIS}{\Rule{LambdaThis}}
\newcommand{\RuleMREFTHIS}{\Rule{MRefThis}}
\newcommand{\RuleSHIFT}[1]{\Rule{{Shift}{#1}}}
\newcommand{\RuleCGEL}{\ensuremath{\textsc{CGE}_\textsc{L}}}
\newcommand{\RuleRETL}{\ensuremath{\textsc{Ret}_\textsc{L}}}
\newcommand{\RuleINSTIMPL}{\Rule{InstImpl}}
\newcommand{\RuleCCALL}{\Rule{CCall}}

\newcommand{\CalledInstanceImplMethod}[3]{\ensuremath{\mathit{InstanceImpl}({#1}, {#2}, {#3})}}

\newcommand{\MH}[2]{\ensuremath{\langle {#1}, {#2} \rangle}}
\newcommand{\MHCGE}[3]{\ensuremath{{#1} \xrightarrow{#3} {#2}}}          
\newcommand{\LambdaCGE}[3]{\ensuremath{{#1} \xrightarrow{#3} {#2}}}  
\newcommand{\MHLookup}[1]{\ensuremath{\mathcal{L}_t}}
\newcommand{\MethodMT}[1]{\ensuremath{\mathit{MethodMT}({#1})}}
\newcommand{\ok}{\ensuremath{\checked}}
\newcommand{\no}{--}
\newcommand{\ReturnVar}[2]{\ensuremath{\mathit{RetVar}({#1}, {#2})}}

\begin{document}

\maketitle

\begin{abstract}
  Java 7 introduced programmable dynamic linking in the form of the
  \invokedynamic{} framework. Static analysis of code containing
  programmable dynamic linking has often been cited as a significant
  source of unsoundness in the analysis of Java programs. For example,
  Java lambdas, introduced in Java 8, are a very popular feature,
  which is, however, resistant to static analysis, since it mixes
  \invokedynamic{} with dynamic code generation. These techniques
  invalidate static analysis assumptions: programmable linking breaks
  reasoning about method resolution while dynamically generated code
  is, by definition, not available statically. In this paper, we show
  that a static analysis can predictively model uses of
  \invokedynamic{} while also cooperating with extra rules to handle
  the runtime code generation of lambdas. Our approach plugs into an
  existing static analysis and helps eliminate all unsoundness in the
  handling of lambdas (including associated features such as method
  references) and generic \invokedynamic{} uses. We evaluate our
  technique on a benchmark suite of our own and on third-party
  benchmarks, uncovering all code previously unreachable due to
  unsoundness, highly efficiently.

\end{abstract}

\section{Introduction}

Object-oriented and functional programming have combined in recent
years to produce hybrid programming languages. Some of these, such as
Scala~\cite{Odersky14}, are new languages, designed from the ground up
to incorporate features of both programming paradigms. Others, for
instance Java~\cite{Gosling2014} and C\#~\cite{Hejlsberg03}, have
adapted to the demand for functional features by carefully adding them
in an existing language design; examples of this evolution are
lambdas~\cite{lambdas} and the streams API~\cite{streams} in the Java
platform and the Language Integrated Query (LINQ) facility in the .NET
ecosystem~\cite{Meijer06}.

On another axis, programming languages occupy different places in the
spectrum between static and dynamic typing. At the extremes,
programming languages either have to supply static (``type'')
information for every entity in the program, or do away with all
such types, in a completely dynamic coding style. In practice, most
programming languages are closer to the middle, having a fundamental
static or dynamic design, while mixing elements from the opposite approach.
For example, the Java Virtual Machine (JVM), the best-established
language runtime system, supports dynamic facilities, such as reflection and
dynamic class loading, that offer significant flexibility, outside the control
of the static type system.

A recent dynamic facility added to the JVM, in order to combine
flexibility with highly optimized performance, is that of programmable
method resolution and dynamic linking, in the form of the
\invokedynamic{} instruction~\cite{Rose09}. The \invokedynamic{}
instruction and its accompanying \texttt{java.lang.invoke} framework
permit the expression of fully dynamic behavior, in much the same way
as traditional Java reflection. However, whereas reflection can be
thought of as dynamically \emph{interpreting} dispatch logic,
programmable linking can be thought of as dynamically \emph{compiling}
dispatch logic, transforming call sites at load time with decisions
possibly cached and subsequently executed at full speed.  This facility enables
the JVM to support dynamic language patterns with great efficiency. As
a result, the framework has also been used to implement Java
lambdas---the newly-added functional feature of the
language.\footnote{We emphasize again that the concepts of lambdas (a
  functional language feature) and programmable linking (a dynamic
  language implementation technique) are orthogonal. Lambdas could be
  implemented via front-end class generation, dynamic code generation
  plus traditional virtual dispatch, or other similar techniques. They
  are implemented using programmable linking in Oracle's JDK only as a
  matter of choice, since the mechanism is flexible, powerful, and
  efficient.}


Dynamic features are welcome by many programmers as they offer a
needed flexibility. However, they come at a cost: static reasoning is greatly
hindered. For instance, static analysis tools for Java are largely
ineffective when faced with \invokedynamic{} code, although static
analysis has long dealt with (statically-typed) dynamic dispatch
(a.k.a. \emph{virtual dispatch}) facilities.
Virtual method resolution in statically-typed bytecode is much easier
to analyze, compared to purely dynamic code that lacks explicit method
signatures. (Virtual dispatch in standard object-oriented languages
performs a dynamic lookup of the function, based on its name,
signature, and the type hierarchy.  This is still significantly
friendlier to static reasoning than completely dynamic calls, of
functions with possibly statically-unknown names or types.)

These problems of static reasoning for the dynamic features of the JVM
(and, by extension, its functional lambdas) have been well
identified. In recent work, \emph{Reif et al.}~\cite{Reif18} and
\emph{Sui et al.}~\cite{Sui18} describe the unsoundness in the
construction of call graphs for Java, caused by features such as
lambdas and \invokedynamic{}. These features are not going away: in a
recent study, Mazinanian~\emph{et al.}~\cite{Mazinanian17} ``found an
increasing trend in the adoption rate of lambdas.'' Also,
Holzinger~\emph{et al.} found method handles, a core part of the
\invokedynamic{} framework, to pose ``a risk to the secure
implementation of the Java platform''~\cite{Holzinger16}. This is a
design problem: to control performance overhead, method handles are
less secure by design, compared to the core reflection
API~\cite{security12}.



In this paper, we propose a static analysis that can successfully
analyze both the \invokedynamic{} framework and its particular
combination with generated code in Java lambdas.
Our analysis cooperates with an existing points-to analysis and
an existing reflection analysis (when needed), in mutually recursive
fashion. The analysis also simulates parts of the Java API that either
do dynamic code generation or call native code, to maintain
soundness. Finally, we supply a special static analysis extension that
can analyze lambdas and method references, without any reflection
support. This last feature permits the static analysis of large Java
code bases without paying the performance overhead of reflection
reasoning.

In more detail, our work makes the following contributions:

\begin{itemize}
\item We offer the first static analysis that handles general-purpose
  \invokedynamic{}---the basis of modern dynamic features of Java. The
  static analysis operates at a deep level: it includes full
  modeling of the underlying
  \texttt{java.lang.invoke} framework: a DSL-like facility for capturing
  and manipulating methods as values.

\item We present a static modeling of Java lambdas---the main
  functional feature of Java. Although lambdas and \invokedynamic{}
  are conceptually orthogonal, in practice lambdas are implemented
  using \invokedynamic{}, making the analyses of the two features
  closely interrelated. Still, the analysis of lambdas is not a mere
  client of the general-purpose \invokedynamic{} analysis, since it
  both needs extra modeling (for generated code) and admits more
  efficient implementation, due to its specialized use of
  \invokedynamic{}, eschewing the need for complex reflection
  reasoning.

\item The analysis is accompanied by a micro-benchmarking suite
  covering many patterns found in realistic uses of lambdas and
  \invokedynamic{}. The suite is independently usable for validation
  of static support of these features.

\item The analysis is evaluated on the third-party suite of \emph{Sui
  et al.}~\cite{Sui18}, which was designed for showcasing the
  unsoundness of call-graph construction under dynamic and functional
  Java features. Our analysis models all general-purpose uses of
  \invokedynamic{} and fully models uses of lambdas.

\end{itemize}

This paper is structured as follows: we first present a set of
examples that explain how the dynamic and functional features of Java
work (Section~\ref{sec:examples}) and proceed to give a more technical
background of these features
(Section~\ref{sec:technical-background}). We then present our
technique for the static analysis of these features, in a declarative
analysis framework (Section~\ref{sec:our-analysis}). We evaluate our
model (Section~\ref{sec:evaluation}), connect with related work
(Section~\ref{sec:related}), and conclude
(Section~\ref{sec:conclusion}).

\section{Motivation and Illustration}
\label{sec:examples}

This section introduces \invokedynamic{} and Java lambdas with the
help of examples.

\subsection{Motivating Example 1: Late Linking}
\label{sec:example-late-linking}

A common use of dynamic linking is for breaking dependencies between
pieces of code so that they do not have to be compiled together. An
example of Java code using \invokedynamic{} to break a compile-time
dependency is shown in Figure~\ref{fig:invokedynamic-example}.
Since Java does not permit \invokedynamic{}-equivalent
expressions at the source level,\footnote{A proposal is underway to allow such
  expressions via intrinsics~\cite{jep303}.} we use in the example an INVOKEDYNAMIC
pseudo-intrinsic that contains the following information:
\begin{itemize}
\item a dynamic name (\texttt{print}),
\item a method type (\texttt{(A)V}),
\item a list of arguments (just \texttt{this.obj} here),
\item a bootstrap method signature (here: \texttt{<A: CallSite
  bootstrap(MethodHandles.Lookup, String, MethodType)>}), and
\item a list of bootstrap arguments (empty in this example).
\end{itemize}

While the code without \invokedynamic{} has to explicitly state which
version to call (and thus store an immutable signature in an
\invokestatic{} in the bytecode), the code using \invokedynamic{}
looks up the method programmatically, via a ``bootstrap'' method, which
initializes the call site. (This lookup could be arbitrarily
complex, although in this example the outcome is always the same.)
Here we note that the programmer could also use classic Java
reflection to do a similar lookup-and-invoke (retrieving a \texttt{Method}
metaobject and calling an \texttt{invoke} method on it), but that would be
inefficient, since standard Java reflection contains an interpretive
layer of introspection. In contrast, \invokedynamic{} can be compiled
away: the bootstrap method is executed at \emph{load time}, not run time
(i.e., not when method \texttt{run} is invoked, but when it is loaded).
The bootstrap method effectively acts as a load-time macro, accepting
as arguments load-time constants (e.g., string constants) or fragments
of uninterpreted expression syntax. This bootstrap method
returns a ``constant call site'', which the JVM can inline in place of
the \invokedynamic{} call as
needed, similar to having the \invokestatic{} call that is missing
from the bytecode.

\begin{figure}
\textbf{Code without invokedynamic}
\begin{lstlisting}[language=Java]
class C implements Runnable {
  A obj;

  C(A obj) {
    this.obj = obj;
  }

  void run() {
    A.print(this.obj);       // Direct call
  }
}

class A {
  public static void print(A a) { }
}

(new C(new A())).run();
\end{lstlisting}
\vspace{-0.5em}
\begin{center}\noindent\rule{0.8\textwidth}{0.4pt}\end{center}
\textbf{Code using invokedynamic}
\begin{lstlisting}[language=Java]
class C implements Runnable {
  A obj;

  C(A obj) {
    this.obj = obj;
  }

  void run() {
    INVOKEDYNAMIC "print" "(A)V" [this.obj]
      <A: CallSite bootstrap(MethodHandles.Lookup,String,MethodType)>
      []
  }
}

class A {
  public static void print(A a) { }
  public static CallSite bootstrap(MethodHandles.Lookup caller,
                                   String name, MethodType type) {
    MethodType mt = MethodType.methodType(Void.TYPE, A.class);
    MethodHandles.Lookup lookup = MethodHandles.lookup();
    MethodHandle handle = lookup.findStatic(A.class, name, mt);
    return new ConstantCallSite(handle);
  }
}

(new C(new A())).run();
\end{lstlisting}
  \caption{Example: using \invokedynamic{} to postpone linking of a method call.}
  \label{fig:invokedynamic-example}
\end{figure}

\subsection{Motivating Example 2: Lambdas}
\label{sec:motivating1}

For a simple program that creates and uses a lambda, we can take the
following example (adapted from the dynamic benchmark of \emph{Sui et
  al.}~\cite{Sui18}):

\begin{lstlisting}[language=Java]
import java.util.function.Consumer;
public class LambdaConsumer {

  public void source() {
    Consumer<String> c = (input) -> target(input);
    c.accept("input");
  }

  public void target(String input) { }

}
\end{lstlisting}

\noindent Here, method \texttt{source()} creates a lambda that
consumes a string value. The lambda takes an \texttt{input} parameter
and calls method \texttt{target()} in its body, passing the parameter
to the callee.

The arrow syntax declares a lambda function, which is rather a mismatch for
object orientation: it looks like a bare method, without an instance
or declaring type. However, that syntax behind the scenes constructs
an object of type \texttt{Consumer}, as shown by the static type of
variable \texttt{c}. This type is one of the ``functional
interfaces''~\cite{Goetz12b} provided by Java, which are interface
types that have a functional flavor, i.e., declare a single method.
Generic typing helps with annotating
uses of such instances (as with the type parameter of
\texttt{Consumer} here).

Indeed, the Consumer type declares a single \texttt{accept} method
that takes a \texttt{String}. Calling that method on a lambda should then
evaluate the body of the lambda with the appropriate parameter passed
to it. If we were to inline the code in the body of \texttt{source()} to
eliminate the lambda, it would read:

\begin{lstlisting}[language=Java]
public void source() {
  target("input");
}
\end{lstlisting}

However, such inlining cannot happen in the general case: lambdas are
often passed to code or returned by it, to be applied in a
location remote to their origin.
Reasoning about the code above is thus based on non-local
(possibly whole-program) reasoning about the
``functional object'' that was created and assigned to variable
\texttt{c}.

In Figure~\ref{fig:invokedynamic-snippet}, we see the bytecode
generated for the two statements in the body of \texttt{source} in our example.
For presentation purposes, instead of
stack-based bytecode, we use the friendlier 3-address Jimple
intermediate language~\cite{Vallee-Rai99b}. (In the Jimple syntax,
the \texttt{invokedynamic} JVM instruction is denoted \texttt{dynamicinvoke}.)
We observe that the call
that generates the lambda does the following:

\begin{itemize}
\item It invokes as a bootstrap method (i.e., the method to execute
  at load-time over the site of \invokedynamic{}) a special
  ``lambda metafactory'' method. Again, this is a method executing
  at load time (i.e., akin to a macro). It processes the call site
  directly and returns a \texttt{CallSite} value, not the 
  \texttt{Consumer} value of the user code.

\item It passes to the lambda metafactory enough information to
  specify what kind of lambda needs to be generated: one with an ``accept''
  method, implementing interface \texttt{Consumer} and capturing from
  its environment parameter \texttt{l0}.
  The \texttt{l0} capture means that the current value of \texttt{this}
    escapes to the new code that will construct the lambda. This is to
    be expected, as the lambda body needs a receiver to resolve the
    call to \texttt{target}.
  \item A method handle pointing to a compiler-generated method
    \texttt{lambda\$source\$0} is also passed as an argument. This
    method encodes the body of the lambda expression. 
\end{itemize}

Note that \texttt{invokedynamic} is used at the site of lambda generation,
not lambda invocation. The latter (in the final line of Figure~\ref{fig:invokedynamic-snippet})
is a regular interface call.

From the perspective of a static analysis, the only method call that
can be resolved in the \invokedynamic{} instruction is the call to the
metafactory but analysis of that cannot complete: the metafactory does
load-time code generation. The compiler-metafactory synergy (of
generating methods at compile time, yet leaving other code generation
and call-site transformation to load time) is a design that cannot be
penetrated by a conventional static analysis. When, in the next
instruction, the static analysis tries to analyze the interface call
on the object returned in the \invokedynamic{} instruction, it cannot
resolve the target method and analysis of this call fails.

\begin{figure}
{\small
\begin{alltt}
l0 := @this: LambdaConsumer;

l1 = dynamicinvoke "accept" <java.util.function.Consumer (LambdaConsumer)>(l0)
       <invoke.LambdaMetafactory: invoke.CallSite metafactory(
         invoke.MethodHandles$Lookup,String,invoke.MethodType,
         invoke.MethodType,invoke.MethodHandle,invoke.MethodType)>
       (class "(Ljava/lang/Object;)V", 
        handle: <LambdaConsumer: void lambda$source$0(String)>,
        class "(Ljava/lang/String;)V");

interfaceinvoke l1.<java.util.function.Consumer: void accept(Object)>("input");
\end{alltt}
}
\caption{The \invokedynamic{} behind a lambda creation (Jimple syntax
  for example in Section~\ref{sec:motivating1}). The package prefix ``\texttt{java.lang}''
  has been removed from all types---e.g., \texttt{invoke.MethodType} is
\texttt{java.lang.invoke.MethodType}.}
\label{fig:invokedynamic-snippet}
\end{figure}

\subsection{Motivating Example 3: Method References}

Java 8 introduced lambdas due to popular demand for the feature but
also because they were needed for scaling stream processing over
multicore hardware~\cite{Goetz12}. Streams were another new functional
feature added to Java, that supported combinator functions over series
of data (``streams''), enabling function composition and higher-order
programming idioms. An example of streams and lambdas is the following
snippet from Urma's streams tutorial~\cite{streams}:

\begin{lstlisting}[language=Java]
List<Integer> transactionsIds = transactions
  .stream()
  .filter(t -> t.getType() == Transaction.GROCERY)
  .sorted(comparing(Transaction::getValue).reversed())
  .map(Transaction::getId)
  .collect(toList());
\end{lstlisting}

Here, we see that function \texttt{filter} takes a lambda using the
arrow syntax. We also see another higher-order feature added in Java
8: method references, such as \texttt{Transaction::getValue} and
\texttt{Transaction::getId}. These pass regular methods as function
parameters to combinator functions \texttt{comparing} and
\texttt{map}.

While the syntax of method references is different compared to
lambdas, these expressions are also implemented by the lambda
metafactory in a similar way. Method references may be a simplified
version of lambdas but they still have semantic complexities as they
can capture a value from the environment for their receiver.

\subsection{Motivating Example 4: SAM Conversion}

The use of lambdas (and, by extension, \texttt{invokedynamic})
in Java is not limited to pure functional
programming patterns. Lambdas are backwards compatible with pre-Java-8 code. In the
following example, we see two \texttt{Runnable} objects being
constructed, both with the same functionality:

\begin{lstlisting}[language=Java]
public class Main {
  public static void main(String[] args) {

    // Use anonymous class.
    Runnable a = new Runnable() {
                   public void run() {
                     System.out.println("Hello.");
                   }
                 };
    a.run();

    // Use a lambda.
    Runnable b = (() -> System.out.println("Hello."));
    b.run();

  }
}
\end{lstlisting}

\noindent The \texttt{Runnable} interface is a standard type of the
Java platform that happens to have a single method. It is, thus, a
``single abstract method'' (``SAM'') type\footnote{Or a ``functional
  interface''~\cite{stateOfTheLambda}.} and the lambda syntax can be
used to generate an instance of it, which can be passed to code
compiled with older Java versions. This approach makes pre-Java-8 code
``forward-compatible to lambdas''~\cite{Goetz12} by viewing all
existing single-method interfaces as lambdas (``SAM
conversion''~\cite{Goetz10,lambdas}). In practice, this ease of
constructing many types as lambdas means that, even in a simple
``hello world'' Java program, several \invokedynamic{} calls
to the lambda metafactory take place.

This has caused a regression in the power of static
analysis tools on bytecode: \emph{unless it supports lambdas, an
  analysis may find fewer facts for the same program under Java 8,
  compared to Java 7}. Java has become more dynamic and functional
under the hood.


\section{Technical Background}
\label{sec:technical-background}

This section gives a basic background on the technology behind method
handles, the \invokedynamic{} framework, lambdas, and method
references. We show as much as needed for the needs of the model of the static
analysis that will follow.

\subsection{Method Handles and Method Types}
\label{sec:backrgound-method-handles}

Two important kinds of values that are used in the rest of this
section are method handles and method types.

\emph{Method handles} are the equivalent of type-safe function
pointers~\cite{Roussel14} and a lightweight alternative to standard
reflective method objects~\cite{Nutter08}. They represent targets for
invocation that can point to methods, constructors, fields, or other
parts of an object~\cite{Roussel14}. There are three basic kinds of
method handles: direct method handles are very similar to pointers;
bound method handles are partial applications of
methods~\cite{hotspot_bmh,Rose11}, and adapter method handles perform
various adjustments of method parameters (e.g., from a flat list of
arguments to a single argument array)~\cite{Rose11}.


\emph{Method types} are type descriptors that help method handle
invocations guarantee run-time type safety. A method type describes the
parameter types and the return type that a method handle can
accept. Method types can be modified to produce new method types: for
example, their return type can be changed and types can be dropped,
changed, or appended~\cite{method_type}.

A method handle can be invoked via two methods called on it:
\begin{itemize}
  \item \texttt{invokeExact()} calls the method handle directly,
    matching its types against the handle method type.
  \item \texttt{invoke()} is more permissive: it permits conversions
    of arguments and return type during the method handle
    invocation. Such conversions must be compatible with appropriate
    conversions of its method type~\cite{method_handle}.
\end{itemize}


The general \texttt{java.lang.invoke} API~\cite{invoke}, offers ways
to compose method handles, convert, fill in, or rearrange their
arguments, perform conditional logic on them, or manipulate them in
other ways. In practice, the method handles API is an embedded
domain-specific language (DSL), which has the flavor of a
combinatorial language over functional types. This DSL does deep
embedding~\cite{Svenningsson15}, i.e. the API creates an intermediate
representation that reflects the semantics of the intended method
handle.

The method handles are translated to an intermediate representation
called \emph{lambda forms}~\cite{lambda_form,deepdive}. (Not to be
confused with the synonymous ``lambda'' high-level functional language feature of the language
that we discuss extensively in this paper.) The lambda form
representations can be cached and reused, interpreted, or compiled
(using Just-in-Time technology). \emph{This aspect of method handles
  argues for a static analysis to model them as primitive concepts:
  since they eventually do dynamic code generation, their semantics
  are impenetrable to a conventional static analysis.}


The compilation of lambda forms creates dynamically-generated bytecode
of a special form, called \emph{anonymous classes}~\cite{anon}. This
is bytecode that is not even visible to the runtime system class
dictionary and is used for fast lightweight code
generation~\cite{Nutter08}. Not only are these classes hidden; they
also violate the read-only invariant of loaded classes in the VM, as
they can patch other classes on the fly.

This design introduces a complete embedded mini-language on top of
bytecode, together with a small implementation (intermediate
representation, interpreter, and compilation back-end). For static
analysis tools to reason about custom dynamic behavior, they must, thus,
reason about this small language, from its front-end API embedding,
through the implementation, to the generated bytecode.

\subsection{The \invokedynamic{} Instruction}
\label{sec:background-invokedynamic}


The JVM was initially used to implement only the Java language. As the
virtual machine became a state-of-the art optimizing Just-in-Time
(JIT) compiler and the underlying platform grew, other
statically-typed object-oriented languages (such as
Scala~\cite{Odersky14} and Fortress~\cite{Allen05}) chose to reuse it
by having a compiler front-end from their syntax to bytecode. At the
same time, the rise of dynamic languages, combined with the desire of
their implementers to reuse the Java platform, led to a proliferation
of dynamic languages implemented on top of the JVM, both existing ones
such as Ruby (JRuby~\cite{Nutter11}), Python
(Jython~\cite{Pedroni02}), and JavaScript
(Rhino/Nashorn~\cite{rhino}), and new ones such as Groovy and
BeanShell. In the meantime, functional features entered the
mainstream, influencing the object-oriented programming paradigm;
functional languages gained enough traction to warrant implementations
on top of the JVM. Examples of functional languages on the JVM are
Clojure~\cite{Halloway09}, the Haskell-inspired Eta~\cite{eta} and
Frege~\cite{frege}, and the Erjang version of
Erlang~\cite{erjang}. Finally, Java itself had to evolve and
incorporate functional features (we describe them in detail in
Section~\ref{sec:background-lambdas}).

To become multi-lingual in an efficient way, the JVM design had to
gain two new powers: the capability to implement diverse dynamic
behaviors; and native support for the basic building block of
functional programming, lambdas. In this subsection, we give an
overview of \invokedynamic{}, while on the next subsection, we will
see how the functional features are supported under the hood as an
instance of dynamic behavior (Section~\ref{sec:background-lambdas}).


A classic characteristic of dynamically-typed languages is their
reliance on runtime optimization for performance, since there are no
statically-available types to use for optimization. Naive
implementations of dynamically-typed languages are slow, since they
are usually interpreters that constantly query metadata to discover
the runtime types of objects in order to perform safe operations on
them. Runtime optimization systems come to the rescue: modern
high-performance dynamic languages profile the running program and
optimize it, often generating good code at runtime, when more
information is known about the behavior of the program (the
``Just-in-Time'' or ``JIT'' approach).

JIT optimization has a long history, for instance one of its
techniques to speed up method calls, ``inline caching'', appears in
the classic implementation of the Smalltalk object-oriented dynamic
language~\cite{Deutsch84}. Today, the JIT approach forms the basic
technology behind successful implementations as diverse as the
cutting-edge Java Virtual Machine~\cite{Lindholm14} or the browser
runtimes of JavaScript that enabled the Web 2.0 wave of applications.

As dynamic languages on the JVM were pushing for more performance on
the JVM, Java 7 introduced a new bytecode opcode,
\invokedynamic{}~\cite{Rose09}, together with an API around it, that
could offer the programmer the capability to completely customize
dynamic program behavior. The program could now implement its own
method dispatch semantics, for example perform linking, unlinking, and
relinking of code on the fly, add or remove fields and methods in
objects, or implement inline caching using plain Java code. The
crucial advantages of this approach, compared to writing adapter code
by hand, are not only in saving engineering effort through a friendly
API, but also in informing the JIT optimizer so that better optimizations
(such as inlining) can happen across dynamic dispatch
borders.

Oracle offers this as motivation: \emph{``The invokedynamic
  instruction simplifies and potentially improves implementations of
  compilers and runtime systems for dynamic languages on the JVM. The
  invokedynamic instruction does this by allowing the language
  implementer to define custom linkage behavior. This contrasts with
  other JVM instructions such as invokevirtual, in which linkage
  behavior specific to Java classes and interfaces is hard-wired by
  the
  JVM.''}\footnote{\url{https://docs.oracle.com/javase/7/docs/technotes/guides/vm/multiple-language-support.html\#invokedynamic}}


Dynamic languages on the JVM were naturally the first users of this
new functionality (JRuby~\cite{first_taste,Wurthinger13},
Jython~\cite{jython_indy}, the Nashorn JavaScript
engine~\cite{laskey11}, Groovy~\cite{groovy}, Redline
Smalltalk~\cite{Nutter14}, and a significant subset of
PHP~\cite{Forax10}), as they could improve their
performance~\cite{Ortin14}. The \invokedynamic{} instruction even
inspired the creation of at least one new JVM-based
language~\cite{Ponge13}. Moreover, this new capability was used for
other applications, such as live code modification~\cite{Ponge12},
aspect-oriented programming~\cite{Nopnipa13}, context-oriented
programming~\cite{Appeltauer10,Maingret15}, multiple
dispatch/multi-methods (a generalization of object-oriented dynamic
dispatch to take more than one method arguments into consideration
when choosing the target method of an invocation)~\cite{Nutter12},
lazy computations~\cite{Nutter12,Goetz11}, generics
specialization~\cite{Goetz16}, implementation of
actors~\cite{Nobakht14}, and dynamically adaptable binary
compatibility via cross-component dynamic linking~\cite{Jezek16}. This
new low-level functionality also became available for programmable
high-level dynamic linking and metaobject protocol implementation via
the Dynalink library~\cite{dynalink}.

Informally, \invokedynamic{} can be seen as configurable
initialization (and possible reconfiguration) of invocations in Java
bytecode. When the JVM loads a class, it resolves every
\invokedynamic{} instruction in it. For every \invokedynamic{} instruction:

\begin{enumerate}
\item A special \emph{bootstrap method} is called. The method reads
  information either embedded in the instruction or coming from the
  constant pool of the class.
\item The bootstrap method returns a \emph{call site} object. That
  object belongs to the instruction location in the bytecode and
  contains a method handle.
\item Since the call site contains a method handle, the invocation is
  resolved now and the call site has been linked. The method handle
  can be thus invoked (see
  Section~\ref{sec:backrgound-method-handles}).
\item The call site is a Java object, so the program can access it and
  can later mutate its method handle so that the invocation is
  effectively re-linked to resolve to another method. This is essential
  for modeling fully dynamic behavior (e.g., making an object support
  an extra method during run time).
\end{enumerate}


The model above means that the program can now control the linking of
method calls. Moreover, this framework makes dynamically-linked
invocations efficient. Since the JVM internally supports
\invokedynamic{}, it can optimize such invocations. For example, if
the call site is a constant call
site,\footnote{\url{https://docs.oracle.com/javase/8/docs/api/java/lang/invoke/ConstantCallSite.html}}
the invocation can be inlined. The efficiency of \invokedynamic{}
invocations has been confirmed by Kaewkasi~\cite{Kaewkasi10} and
Ortin~\emph{et al.}~\cite{Ortin14}.

\subsection{Method References and Lambdas}\label{sec:background-lambdas}


As seen in the examples of Section~\ref{sec:examples}, method
references and lambdas are functional programming features added to
Java for more expressive power. Eventually, Java 8 implemented these
two features with \invokedynamic{}~\cite{Goetz11}. A crucial
motivation for this implementation choice has been compatibility,
i.e., to avoiding a commitment to a single bytecode-visible
implementation of lambdas (e.g., as classes). Describing the
implementation of lambdas in terms of \invokedynamic{} gives the Java
compiler developers the freedom to later change the underlying
implementation, without breaking binary
compatibility~\cite{Goetz11,Goetz12}. The only trace of the
translation of lambdas inside the bytecode is an \invokedynamic{}
call to a specific lambda metafactory, but the code emitted by that
may later change.

Both lambdas and method references use the same implementation
technique: \invokedynamic{} sites that use special bootstrap methods,
the lambda metafactories~\cite{metafactory}. A lambda metafactory
initializes a call site so that it contains a lambda factory, i.e., it
can generate functional objects. The Java 8 lambda metafactory
generates an inner class that implements the functional
interface.\footnote{\url{https://bugs.openjdk.java.net/browse/JDK-8000806}}
The functional objects created can either be stateless or access
values from their enclosing environment~\cite{lambdaTranslation}. The
implementation of lambdas is a thin layer of code that only uses small
pieces of dynamically-generated code as glue.

In practice, the Java compiler creates appropriate methods for the
bodies of lambdas (\emph{implementing methods}) and registers method
handles of them in the constant pool. These method handles are then
used in \invokedynamic{} invocations to the lambda metafactories,
together with any values captured from the environment. The lambda
metafactories can then create new anonymous classes that can be used
to instantiate the functional objects and forward method calls to the
implementing methods.

\section{Static Analysis}
\label{sec:our-analysis}

We next present our model for handling the \texttt{java.lang.invoke}
API (i.e., method handles), the \invokedynamic{} instruction (in
general), as well as Java lambdas. We offer a declarative set of
inference rules that appeal to relations defined and used by an
underlying value-flow/points-to static analysis. Our implementation is
on the declarative Doop framework~\cite{oopsla/BravenboerS09}, so it
is to a great extent isomorphic to the analysis model presented.

The essence of our analysis approach is threefold:

\begin{itemize}
\item Our baseline model gives semantics to method handles. (This is also
  the main novelty of our approach: the deep modeling of the \texttt{java.lang.invoke}
  API at its most fundamental level.) This
  requires appealing to an existing value flow analysis, since method
  handles have no hard-coded signatures in the bytecode: they offer
  \texttt{invoke} operations that are ``signature-polymorphic''. Therefore,
  any resolution of method handles requires a static model of all possible
  signature arguments to \texttt{invoke} instructions.
  Modeling the semantics of method handles is necessary since their
  implementation is un-analyzable, relying on run-time code generation
  (via the aforementioned ``lambda forms''). Furthermore, this model
  requires static analysis of Java reflection, since method handles
  can also be looked up via reflection operations (e.g., by method types
  generated via reflective class values, or by ``unreflecting''
  method objects into method handles).

\item Based on the modeling of method handles, we straightforwardly
  model \invokedynamic{} as an invocation of a method handle computed
  by a bootstrap method.

\item Reasoning about lambdas appeals to a part of the \invokedynamic{} reasoning.
  However, modeling lambdas both requires extra reasoning (because of dynamic
  code generation) and can avoid the need for expensive reflection analysis, since
  the method handles computed for lambdas do not employ reflection.
\end{itemize}

\subsection{Model Basics}

We assume the following domains and (meta)variables, also listing
some simple convenience predicates along the way:

\begin{itemize}
\item $s \in S$ are strings.
\item $n, k \in \mathcal{N}$ are numbers.
\item The symbol $*$ denotes arguments that can be ignored.
\item $v \in V$ are variables, $\val{} \in \mathit{Val}$ are values,
  $\lambda{} \in \mathit{Val}$ are functional objects.
\item $t \in T$ are types while $\intf{} \in \mathit{T^I} \subset T$
  are interface types. Constructor \mockC{t}{i} creates a mock object
  of type $t$ that corresponds to an instruction $i$. The Class
  metaobject of a type $t$ is given by \ReifiedClass{t} and is a value.
\item $m \in M$ are methods. The formal of $m$ at position $n$ is
  represented as $\Frm{m}{n}$. The special ``this'' variable of an
  instance method $m$ is represented as \ThisVar{m}. The Method
  metaobject of a method $m$ is given by \ReifiedClass{m} and is a
  value. We use the following predicates:
  \begin{itemize}
  \item \IsConstructor{m}: $m$ is a constructor method.
  \item \IsStatic{m}: $m$ is a static method.
  \item $m \in t$: $m$ is declared in type $t$.
  \item The return variable $v$ of $m$ is represented as
    \ReturnVar{v}{m}.
  \end{itemize}
\item $m_t = \{t, [ t_0, \ldots, t_{n-1} ] \} \in M^T, n \ge 0$ are
  method types, which are pairs of a return type $t$ and a (possibly
  empty) list of parameter types. Predicate \AsType{m_t^1}{m_t^2}
  holds when method type $m_t^2$ has the same arity as $m_t^1$, and
  for every pair $t, t'$ of $m_t^1$ and $m_t^2$ (at the same
  position), it holds that the two types are compatible:
  \TypeCompat{t}{t'}. (\TypeCompat{t}{t'} is one of the analysis's
  main input predicates from Figure~\ref{fig:relations}.)
  This type compatibility represents the \texttt{asType} rules of the
  specification~\cite{method_handle}.
  Function \MethodMT{m} maps a method $m$ to its method type.
\item $i \in I$ are invocation instructions. Predicate $i \in m$ means
  that instruction $i$ belongs to method $m$. The actual parameter
  that is passed at invocation $i$ in position $n$ is represented as
  \Act{i}{n}. For \invokedynamic{} instructions, these are the
  non-bootstrap parameters of the bytecode instruction. If instruction
  $i$ returns a value, \AssignRet{i} is the variable that will hold
  the returned value.
\item $h \in \mathit{MH}$ are method handles. A method handle $h$ has
  the form \MH{m}{m_t}, which is a pair of a method $m$ and a method
  type $m_t$. We also assume predicate $\DMHLookup{t}{s}{m_t}$, which
  returns the direct method handle that corresponds to a method with
  name $s$, declared in type $t$, with method type $m_t$.
  Constructor \mockH{t}{h} creates a mock object
  of type $t$ that corresponds to method handle $h$.
\item $c \in C$ are call site identifiers. (These are different from
  mere instructions: because of the dynamic nature of calls, the same
  instruction can play the role of distinct call sites.)
\item We assume lookup objects \MHLookup{t}, one for each type
  $t$. These are opaque objects in the \texttt{java.lang.invoke} API
  that are used as intermediate values in a lookup: to retrieve, e.g.,
  a method handle, first one retrieves a lookup object over a type, and
  subsequently uses it with method-identifying information.\footnote{Maintaining
    a distinct lookup object for each type also shows that our
    technique can potentially track access restrictions per type, as
    mandated by the specification of method lookup
    objects: \url{https://docs.oracle.com/javase/8/docs/api/java/lang/invoke/MethodHandles.Lookup.html} .}
\end{itemize}

The table in Figure~\ref{fig:relations} lists the main relations that will
be used in the analysis rules (i.e., all relations other than convenience predicates described
earlier). We annotate each relation with
\In{} if it is consumed by our rules and \Out{} if our rules inform
it. Relation \VPT{v}{\val} is both \In{} and \Out{}, since our analysis
is mutually recursive with the existing points-to analysis. Relations
annotated with \Internal{} are intermediate relations used in the
analysis, that may not be externalized.

\begin{figure}
  \begin{tabular}{|lll|}
    \hline
    \textbf{Relation}            & \textbf{Description}                                                & \textbf{Use}  \\
    \hline
    \VPT{v}{\val}                & Variable $v$ points to $\val$.                                      & \In{}, \Out{} \\
    \IFPT{f}{\val}               & Field $f$ points to $\val$.                                         & \In{}         \\
    \APT{v}{\val}                & Variable $v$ is an array and $v[i]$ points to $\val$ for some $i$.  & \Out{}        \\
    \MHCGE{i}{m}{h}              & Instruction $i$ calls method $m$ using method handle $h$.           & \Out{}        \\
    \LambdaCGE{i}{m}{\lambda}    & Instruction $i$ calls method $m$ using functional object $\lambda$. & \Out{}        \\
    \TypeCompat{t}{t'}           & Types $t$ and $t'$ are either subtypes of each other or can         & \In{}         \\
                                 & be converted to each other via boxing or unboxing.                  &               \\
    \InvokedynamicCallSite{c}{i}{t}
                                 & Instruction $i$ creates call site $c$ with                          & \Internal{}   \\
                                 & dynamic return type $t$.                                            &               \\
    \CallSiteContents{c}{h}{m}   & Call site $c$ contains meth. handle $h$ pointing to                 & \Internal{}   \\
                                 &  method $m$.                                                        &               \\
    \LambdaMetafactoryInvoke{i}{s}{\intf} & Lambda metafactory invocation at instruction $i$, with     & \Internal{}   \\
                                 & dynamic method name $s$ and functional interface $\intf$.           &               \\
    \LambdaObject{\lambda}{m}{s}{i} & Functional object $\lambda{}$ with implementing method $m$,      & \Internal{} \\
                                 & dynamic method name $s$ and \invokedynamic{}                        &               \\
                                 & source instruction $i$.                                             &               \\
    \LambdaCaptured{i}{n}{\val}  & Instruction $i$ captures environment value $\val$ at                & \Internal{}   \\
                                 & position $n$.                                                       &               \\
    \CalledInstanceImplMethod{i}{m}{\lambda} & Functional object $\lambda$, generated at instruction $i$, uses & \Internal{}\\
                                 & non-static method $m$ as implementing method.                       &               \\
    \hline
  \end{tabular}
  \caption{Analysis relations.}
  \label{fig:relations}
\end{figure}

\subsection{Model: Method Types and Method Handles}

We show how the analysis can understand the APIs of method types and
method handles. This includes handling the \emph{polymorphic
  signatures} of Java bytecode.

A fundamental problem in the static analysis of method handles is that
they contain native code, for example their ``invoke'' methods that
must be used to do the method call are native.

The basic relation in this model is \MHCGE{i}{m}{h} which is a
call-graph edge from instruction $i$ to method $m$ annotated with a
method handle $h$. This relation is both created by rules (that
discover method handle invocations) and consumed by rules (that handle
argument passing and value returns).

The method handle invocation rules are shown in
Figure~\ref{fig:method-handles-invocation} while
Figure~\ref{fig:method-handles-api} shows the rules that simulate part
of the method handles API. For clarity, we omit packages from
qualified types (e.g., we write \texttt{MethodHandle} instead of
\texttt{java.lang.invoke.MethodHandle}).

The rules of Figure~\ref{fig:method-handles-invocation} are relatively
straightforward, capturing regular calling semantics for
method handle invocations, once a method handle value has been determined.
Interesting elements include the mutual recursion with an existing
points-to analysis, as well as the construction of new (mock) objects,
per the API specification, when a method handle that corresponds to
a constructor is invoked.

\begin{description}
\item [Rule \RuleMHMETHOD.] This rule creates a method handle $h$
  and a method type $m_t$ for every method found in the program.


\item[Rule \RuleMHCGE.]  This rule informs the method handles call
  graph relation that an invocation $i$ calls method $m$ using method
  handle $h$ (notation: \MHCGE{i}{m}{h}).


\item[Rules \RuleRETH{} and \RuleMHARGS.]  These rules pass
  arguments and return parameters.

\item[Rule \RuleMHCONSTR.] For method handles that correspond to
  constructors, a mock value is constructed and both the \texttt{this}
  variable in it and the return value of the invocation point to this
  value.
\end{description}

\noindent The rules of Figure~\ref{fig:method-handles-api} are a bit
more demanding, since they capture precisely the semantics of the
\texttt{java.lang.invoke} API, including lookup objects, using
reflection to retrieve method handles, and more.

\begin{description}
\item[Rule \RuleASTYPE.] This rule models the \texttt{asType()} method of
  the \texttt{MethodHandle} API using predicate \AsType{m_t^1}{m_t^2}.

\item[Rule \RuleMHLOOKUP.] This rule models the per-type lookup
  object needed to find method handles. The \texttt{lookup()} method
  modeled in this rule is caller-sensitive~\cite{method_handles}, thus
  the caller type $t$ characterizes the returned lookup object and is
  available for future uses of the object.

\item[Rule \RuleMHLOOKUPC.] This rule models the connection between
  a lookup object and its type (e.g., to be used in the code for
  accessibility checks).

\item[Rule \RuleUNREFLECT.] This rule models the API methods that
  bridge the Reflection API with the \texttt{java.lang.invoke}
  API. These methods convert reified methods/constructors to method
  handles.

\item[Rule \RuleFIND.] This rule models the API methods that look up a
  virtual or static method via a lookup object, returning a method
  handle.

\item[Rule \RuleMTYPE.] This rule models the two-argument method
  \texttt{methodType()} of class \texttt{MethodHandle}. The other
  overloaded versions of \texttt{methodType()} are modeled
  similarly. The rule needs access to reflection support, since it
  takes advantage of points-to information that points to reified
  Class objects.

\end{description}

\begin{figure}
  \begin{centering}

  \begin{tabular}{c}
    \infer[\RuleMHMETHOD]{
      \MH{m}{m_t}
    }{
      m_t = \MethodMT{m}
    } \\
  \end{tabular}\\[1em]


  \begin{tabular}{c}
    \infer[\RuleMHCGE]{
      \MHCGE{i}{m}{h} 
    }{
      i = v.\texttt{<MethodHandle.invokeExact>($\ldots$)} \ruleAnd{} 
      \VPT{v}{h} \ruleAnd{}
      h = \MH{m}{*}
    }
  \end{tabular}\\[1em]

  \begin{tabular}{cc}
    \infer[\RuleMHARGS]{
      \VPT{\Frm{m}{k}}{\val}
    }{
      \MHCGE{i}{m}{h} \ruleAnd{}
      \VPT{\Act{i}{n}}{\val}
    } &
    \infer[\RuleRETH]{
      \VPT{v'}{\val}
    }{
      \MHCGE{i}{m}{h} \ruleAnd{}
      \ReturnVar{v}{m} \ruleAnd{}
      \VPT{v}{\val} \ruleAnd{}
      \AssignRet{i} = v'
    }
  \end{tabular}\\[1em]



  \begin{tabular}{c}
    \infer[\RuleMHCONSTR]{
      \VPT{\ThisVar{m}}{\val} \ruleAnd{}
      \VPT{v}{\val}
    }{
      \MHCGE{i}{m}{h} \ruleAnd{}
      \IsConstructor{m} \ruleAnd{}
      \val = \mockH{t}{h} \ruleAnd{}
      \AssignRet{i} = v
    }
  \end{tabular}\\[1em]

  \end{centering}
  \caption{Rules for handling method handle invocations.}
  \label{fig:method-handles-invocation}
\end{figure}

\begin{figure}
  \begin{centering}
  \begin{tabular}{c}
    \infer[\RuleASTYPE]{
      \VPT{v''}{\MH{m}{m_t^2}}
    }{
      i = v.\texttt{<MethodHandle.asType>($v'$)} \ruleAnd{}
      \VPT{v}{\MH{m}{m_t^1}} \\ 
      \VPT{v'}{m_t^2} \ruleAnd{}
      \AsType{m_t^1}{m_t^2} \ruleAnd{}
      \AssignRet{i} = v''
    }
  \end{tabular}\\[1em]

  \begin{tabular}{c}
    \infer[\RuleMHLOOKUP]{
      \VPT{v}{\MHLookup{t}}
    }{
      i = \texttt{<MethodHandles.lookup>()} \ruleAnd{}
      i \in m \ruleAnd{}
      m \in t \ruleAnd{}
      \AssignRet{i} = v
    }
  \end{tabular}\\[1em]

  \begin{tabular}{c}
    \infer[\RuleMHLOOKUPC]{
      \VPT{v'}{\ReifiedClass{t}}
    }{
      i = v.\texttt{<MethodHandles.Lookup.lookupClass>()} \ruleAnd{}
      \VPT{v}{\MHLookup{t}} \ruleAnd{}
      \AssignRet{i} = v'
    }
  \end{tabular}\\[1em]

  \begin{tabular}{c}
    \infer[\RuleUNREFLECT]{
      \VPT{v'}{\MH{m}{m_t}}
    }{
      \AssignRet{i} = v' \ruleAnd{}
      \MethodMT{m} = m_t \ruleAnd{}
      i = \texttt{<MethodHandles.Lookup.$s$>($v$)} \\
      \VPT{v}{\ReifiedMethod{m}} \ruleAnd{}
      s \in \{ \texttt{unreflect}, \texttt{unreflectSpecial}, \texttt{unreflectConstructor} \}
    }
  \end{tabular}\\[1em]

  \begin{tabular}{c}
    \infer[\RuleFIND]{
      \VPT{v'}{h}
    }{
      i = v.\texttt{<MethodHandles.Lookup.$s$>($v_0$, $v_1$, $v_2$)} \ruleAnd{}
      s \in \{ \texttt{findVirtual}, \texttt{findStatic} \} \ruleAnd{}
      \VPT{v}{\MHLookup{t}} \\
      \AssignRet{i} = v' \ruleAnd{}\ruleAnd{}
      \VPT{v_0}{\ReifiedClass{t'}} \ruleAnd{}
      \VPT{v_1}{s} \ruleAnd{}
      \VPT{v_2}{m_t} \ruleAnd{}
      \DMHLookup{t'}{s}{m_t} = h
    }
  \end{tabular}\\[1em]

  \begin{tabular}{c}
    \infer[\RuleMTYPE]{
      \VPT{v}{\{ t_0, [ t_1 ]\}}
    }{
      i = v.\texttt{<MethodType.methodType>($v_0$, $v_1$)} \hspace{1.5cm} \\ 
      \VPT{v_0}{\ReifiedClass{t_0}} \ruleAnd{}
      \VPT{v_1}{\ReifiedClass{t_1}} \ruleAnd{}
      \AssignRet{i} = v
    }
  \end{tabular}\\[1em]

  \end{centering}
  \caption{Rules for handling part of the method handles API.}
  \label{fig:method-handles-api}
\end{figure}

\subparagraph*{Reflection Support.}
A useful subset of these rules does not need reflection support in the
analysis. For some programs, method types and method handles may come
from the constant pool instead of being looked up by the
\texttt{java.lang.invoke} API; for such code, our rules do not require
reflection support.


\subparagraph*{\texttt{invoke()} vs. \texttt{invokeExact()}.}
As mentioned in Section~\ref{sec:backrgound-method-handles}, the
method handle API offers two different ways to invoke a method
handle. The most fundamental is \texttt{invokeExact()}, which assumes
the arguments and the return value have types that exactly match the
method type of the method handle. In contrast, \texttt{invoke()}
permits conversions in arguments and return values, as if the method
handle could successfully change its method type via the
\texttt{asType()} method. For presentation purposes, we only show the
rules for \texttt{invokeExact} in
Figure~\ref{fig:method-handles-invocation} and the rules for
\texttt{asType()} in Figure~\ref{fig:method-handles-api}. The handling
of \texttt{invoke()} follows directly from these rules, accounting for
autoboxing in the case of primitive conversions. The handling of
\invokedynamic{} (shown in Section~\ref{sec:generic-invokedynamic}) is
not affected, since that only needs the functionality of invocations
via \texttt{invokeExact}~\cite{Lindholm14}.

\subparagraph*{Generalized method handles.}
Method handles are also able to represent fields; we don't model this
behavior here since it is not important for the \invokedynamic{}
analysis (that follows in the next section) but it is a simple
extension of our model.


\subsection{Generic Handling of \invokedynamic{}}
\label{sec:generic-invokedynamic}

We next discuss the static modeling of \invokedynamic{}
instructions. The model effects the dynamic linking that eventually
computes a method handle and invokes it. The key concept employed is
call sites ($c \in C$). These are the return objects of
\invokedynamic{} bootstrap methods (as determined by regular points-to
analysis) and internally use method handles to determine the calling
behavior.

Our rules model the \invokedynamic{} framework in order to discover
the method handles contained in each call site. When a method handle
$h$ that maps to a method $m$ is discovered to be contained in the
call site of instruction $i$, a new call-graph edge \MHCGE{i}{m}{h} is
created and the rules of the previous section analyze the method
handle invocation. The rules for handling \invokedynamic{} invocations
are shown in Figure~\ref{fig:generic-invokedynamic}. Evaluation-wise,
these rules \emph{precede} the earlier rules that give semantics to
method handles: The purpose of the rules in
Figure~\ref{fig:generic-invokedynamic} is to express what an
\invokedynamic{} does in terms of method handles, so that the earlier
reasoning can take over.

We extend the earlier domains and predicates with:

\begin{itemize}
\item $I^d \subset I$ are \invokedynamic{} instructions.  Predicate
  \CGEB{i}{m} holds when an \invokedynamic{} instruction $i$ calls
  bootstrap method $m$.\footnote{We assume that all \invokedynamic{}
    instructions call their bootstrap methods when their containing
    type is loaded.} We also assume the following \invokedynamic{}
  projections:
  \begin{itemize}
  \item $\InvokedynamicBoot{} : I^d \rightarrow M$ returns the bootstrap method.
  \item $B^p : I^d \rightarrow n \rightarrow V$ returns the bootstrap parameter at position $n$.
  \item $\Dyn{} : I^d \rightarrow (S \times{} M^T)$ returns the
    dynamic method name / method type pair.
  \end{itemize}
\end{itemize}

The rules are explained below:

\begin{description}


\item[Rules \RuleBARGS, \RuleBARGSZ, and \RuleBARGSV.] The first
  rule passes arguments to the boot method, shifted by three
  positions, since the first three arguments are filled in by the JVM
  (and handled by rule \RuleBARGSZ). Boot methods such as the
  alt metafactory may also take varargs that require special handling
  by the JVM, thus we also have rule \RuleBARGSV. Note the
  introduction of an artificial (mock) array object to maintain
  the vararg values.

\item[Rule \RuleRETB.] This is the standard rule that returns value
  from a method call. It is adapted here for completeness, for the
  case of bootstrap method invocations.

\item[Rule \RuleCSITE.] This rule stores information about a call
  site object computed at an \invokedynamic{} instruction.

\item[Rules \RuleCALLSITECONTENTS{1} and \RuleCALLSITECONTENTS{2}.]
  These two rules relate a call site object with its method handle and
  the method it points to.

\item[Rule \RuleMHCGEI.] This rule relates the \invokedynamic{} call
  site and its method handle to create call-graph edges with method
  handle semantics. From this point on, the rules in the previous
  section take over and complete the method handle invocation.

\end{description}

\subparagraph*{Reflection Support.}
The rules presented in the subsection do not require reflection. For
example, a program which contains an \invokedynamic{} instruction that
passes a method handle constant (read from the class constant pool) to
its bootstrap method, can be analyzed without reflection support. In
practice, however, bootstrap methods often employ reflective reasoning
to compute the method handle that will be returned in the call site
return value, and thus reflection support should be provided.

\begin{figure}
  \begin{centering}
    \begin{tabular}{cc}
      \infer[\RuleBARGSZ]{
        \VPT{\Frm{m}{0}}{\MHLookup{t}} \ruleAnd{}
        \VPT{\Frm{m}{1}}{s} \ruleAnd{}
        \VPT{\Frm{m}{2}}{m_t}
      }{
        \CGEB{i}{m} \ruleAnd{}
        \Dyn(i) = \langle s, m_t \rangle
      } &
      \infer[\RuleBARGS]{
        \VPT{\Frm{m}{n + 3}}{\val}
      }{
        \CGEB{i}{m} \ruleAnd{}
        \VPT{B^p(i, n)}{\val}
      }
    \end{tabular}\\[1em]

    \begin{tabular}{c}
      \infer[\RuleBARGSV]{
        \VPT{\Frm{m}{3}}{a} \ruleAnd{}
        \APT{\val'}{\val}
      }{
        \CGEB{i}{m} \ruleAnd{}
        \val' = \mockC{\texttt{java.lang.Object[]}}{i} \ruleAnd{}
        \VPT{B^p(i, n)}{\val} \ruleAnd
        n > 2
      }
    \end{tabular}\\[1em]

    \begin{tabular}{c}
    \infer[\RuleRETB]{
      \VPT{v'}{\val}
    }{
      \CGEB{i}{m} \ruleAnd{}
      \ReturnVar{v}{m} \ruleAnd{}
      \VPT{v}{\val} \ruleAnd{}
      \AssignRet{i} = v'
    }
    \end{tabular}\\[1em]
      
    \begin{tabular}{c}
      \infer[\RuleCSITE]{
        \InvokedynamicCallSite{c}{i}{t}
      }{
        \Dyn(i) = \langle *, m_t \rangle \ruleAnd{}
        m_t = \{ t, * \} \ruleAnd{}
        \InvokedynamicBoot(i) = m \ruleAnd{}
        \ReturnVar{v}{m}  \ruleAnd{}
        \VPT{v}{c}
      }
    \end{tabular}\\[1em]

    \begin{tabular}{c}
      \infer[\RuleCALLSITECONTENTS{1}]{
        \CallSiteContents{c}{h}{m}
      }{
        \InvokedynamicCallSite{c}{*}{t} \ruleAnd{}
        \IFPT{\Target{c}}{h} \ruleAnd{}
        h = \MH{m}{\{ t, * \}}
      }
    \end{tabular}\\[1em]

    \begin{tabular}{c}
      \infer[\RuleCALLSITECONTENTS{2}]{
        \CallSiteContents{c}{h}{m}
      }{
        \InvokedynamicCallSite{c}{*}{t} \ruleAnd{}
        \IFPT{\Target{c}}{h} \ruleAnd{}
        h = \MH{m}{*} \ruleAnd{}
        \IsConstructor{m} \ruleAnd{}
        m \in t
      }
    \end{tabular}\\[1em]

    \begin{tabular}{c}
      \infer[\RuleMHCGEI]{
        \MHCGE{i}{m}{h}
      }{
        \InvokedynamicCallSite{c}{i}{t} \ruleAnd{}
        \CallSiteContents{c}{h}{m} \ruleAnd{}
        h = \MH{*}{\{ t, * \}}
      }
    \end{tabular}\\[1em]
  \end{centering}
\caption{Rules for generic handling of \invokedynamic{}.}
\label{fig:generic-invokedynamic}
\end{figure}

\subsection{Model: Method References and Lambdas}

Both method reference expression and lambdas are implemented by the same
machinery, a ``lambda metafactory''~\cite{metafactory}. At a very high
level, the metafactory takes two arguments, (1)~a method handle
pointing to a method $m$ and (2)~a SAM type $t$, and returns a
functional object implementing $t$ whose (single) method calls
$m$.

The functional object may be an instance of a new
dynamically-generated class, thus a naive points-to analysis cannot
penetrate the object to analyze calls on it. Our analysis understands
the semantics of the functional objects created by the dynamic linking
and method resolution of the metafactory, and creates a mock value in
place of the functional object. That value can be propagated in the
program as usual by the underlying points-to analysis. Appropriate
metadata on the value help the analysis compute intended semantics
such as the invocation target or the captured values of the
environment.

\subparagraph*{The Three Phases of \invokedynamic{} for Lambdas.}
When used for lambdas, functional object creation by the lambda
metafactory works in three phases~\cite{metafactory}:
\begin{enumerate}
\item \textbf{Linkage.} The bootstrap method is called and a call site
  object is returned, at the location of the \invokedynamic{}
  instruction. The bootstrap method being the ``metafactory'', the
  call site is then a ``lambda factory'', which must be invoked to
  produce a functional object.
\item \textbf{Capture.} The method handle in the call site object is
  invoked, possibly with some arguments. This permits different
  behavior for different contexts by capturing values of the enclosing
  environment. The result is the functional object.
\item \textbf{Invocation.} The functional object can then be passed
  around in the code and the method of its functional interface can be
  eventually called.
\end{enumerate}

The rules that enable analysis of method references are shown in
Figure~\ref{fig:lambdas}. The basic idea is to create mock values in
the analysis for functional objects and simulate all three phases so
that calls are correctly resolved. We assume the following domains,
(meta)variables, and predicates:

\begin{itemize}
\item The \LambdaMetas{} constant stands for the lambda metafactory~\cite{metafactory} of the
  OpenJDK.


\item $\lambda \in \mathit{Val}$ ranges over functional objects.
\item \Arity{i} returns the arity of instruction $i$ (the number of
  actual parameters passed to the functional object).
\end{itemize}

The rules are explained below:

\begin{description}

\item[Rule \RuleMETAFACTORY.] This rule marks an \invokedynamic{}
  invocation as a lambda metafactory invocation.

\item[Rule \RuleLAMBDA.] This rule creates the mock functional
  object $\lambda$ that will propagate in the program and behave (in
  the analysis) as if it was an object created by the metafactory. The
  object keeps related metadata in relation
  \LambdaObject{\lambda}{m}{s}{i}: its implementing method $m$ (found
  in a constant method handle argument of the metafactory), the name
  of the functional interface method it implements, and the
  \invokedynamic{} instruction $i$ that created the functional object.

\item[Rule \RuleCAPTURE.] This rule records possibly captured values from the
  enclosing environment. (All arguments are eagerly recorded as possible captured
  values and the appropriate capture arguments are recognized in later rules
  \RuleCAPTARGS and \RuleLAMBDATHIS.)

\item[Rule \RuleCGEL.] This rule creates call-graph edges to the
  actual implementing method of the functional object. Following these
  edges bypasses the dynamically-generated classes and lets the static
  analysis discover the code of method references and lambdas.

\item[Rule \RuleRETL.] This is the standard rule for return values
  from methods.

\item[Rule \RuleINSTIMPL.] This rule records that a functional object
  is implemented by a non-static method. This means that further rules
  should discover the receiver and pass it to the method.


\item[Rules \RuleSHIFT{1}, \RuleSHIFT{2}, and \RuleSHIFT{3}.] These
  rules populate relation \ParamsReceiverShiftRight{\lambda}{m}{n}{k},
  which records if the arguments passed to the functional object must
  be shifted to make room for a receiver. This is because instance
  methods may implicitly consume one of the actual arguments of the
  \invokedynamic{} or of the functional interface invocation, to use
  as the receiver. Static methods take all \invokedynamic{}-actual
  arguments before the ones passed to the functional object during
  method invocation.

\item[Rule \RuleLARGS.] This passes arguments to the implementing
  method, from the method invocation on the functional object. The
  shifting of parameters addresses a number of patterns that the
  metafactory follows to capture and pass values from the environment.

\item[Rule \RuleCAPTARGS.] This rule passes captured arguments to
  the implementing method.

\item[Rule \RuleLAMBDATHIS.] This rule handles the pattern of
  captured receiver parameters.

\item[Rule \RuleMREFTHIS.] This rule handles the pattern where a
  method reference to an instance method has not captured a receiver,
  but will receive it during invocation as an extra argument.



\item[Rule \RuleCCALL.]  This handles the special case where a
  method reference points to a constructor (is thus a ``constructor
  reference''). Since constructor methods are \texttt{void} and assume
  an already constructed (but not initialized) object, this rule
  creates such an object and binds it both to the 'this' variable of
  the constructor and the return variable of the invocation.

\end{description}

\subparagraph*{Additional Features.}
The JDK also has a second metafactory, the ``alt metafactory'': a
generalization of the lambda metafactory
that provides additional features, such as bridging, support for
multiple interfaces, and serializability. We do not model such extra
properties of its lambdas here, but these features are type-based so
they are amenabe to handling in a similar way to the rules we already
present.

\subparagraph*{Reflection Support.}
The method handles passed to the metafactory are statically known:
either the programmer provided them as method references or the
compiler generated them for lambdas. Thus our rules for handling
lambdas and method references do not need reflection support; the only
method handles used come from the constant pool. This means that our
approach can integrate with the baseline configuration of a points-to
analysis, in order to analyze programs without overhead due to
reflection support.

\subparagraph*{Context sensitivity.}
The analysis, as presented, has a context-insensitive formulation, to
avoid unnecessary complication of the rules.  Careful (but
conceptually standard) addition of context elements to predicates (as
shown, e.g., in reference \cite{PGL-014}) produces a context-sensitive
version.  Our implementation is fully context sensitive.






\begin{figure}
  \begin{centering}
  \textbf{LINKAGE}\\[1em]

  \begin{tabular}{c}
    \infer[\RuleMETAFACTORY]{%
      \LambdaMetafactoryInvoke{i}{s}{\intf}
    }{
      \CGEB{i}{m} \ruleAnd
      m \in \LambdaMetas{} \ruleAnd{}
      \Dyn(i) = \langle s, m_t \rangle \ruleAnd{}
      m_t = \{ \intf, * \}
    }
  \end{tabular}\\[1em]

  \begin{tabular}{c}
    \infer[\RuleLAMBDA]{%
      \VPT{v}{\lambda} \ruleAnd{}
      \LambdaObject{\lambda}{m}{s}{i}
    }{
      \LambdaMetafactoryInvoke{i}{s}{\intf} \ruleAnd{}
      \VPT{B^p(i, 1)}{\MH{m}{*}} \ruleAnd{}
      \AssignRet{i} = v \ruleAnd{}
      \lambda = \mockC{\intf}{i}
    }
  \end{tabular}\\[1em]

  \vspace{1em}
  \textbf{CAPTURE}\\[1em]

  \begin{tabular}{c}
    \infer[\RuleCAPTURE]{%
      \LambdaCaptured{i}{n}{\val}
    }{
      \LambdaMetafactoryInvoke{i}{*}{*} \ruleAnd{}
      \VPT{\Act{i}{n}}{\val}
    }
  \end{tabular}\\[1em]

  \vspace{1em}
  \textbf{INVOCATION}\\[1em]

  \begin{tabular}{c}
    \infer[\RuleCGEL]{%
      \LambdaCGE{i}{m}{\lambda}
    }{
      \LambdaObject{\lambda}{m}{s}{*} \ruleAnd{}
      \VPT{v}{\lambda} \ruleAnd{}
      i = v.\texttt{<$s$>($\ldots$)}
    }
  \end{tabular}\\[1em]

  \begin{tabular}{c}
    \infer[\RuleRETL]{
      \VPT{v'}{\val}
    }{
      \LambdaCGE{i}{m}{\lambda} \ruleAnd{}
      \Ret{m}{n}{v} \ruleAnd{}
      \VPT{v}{\val} \ruleAnd{}
      \AssignRet{i} = v'
    }
  \end{tabular}\\[1em]

  \begin{tabular}{c}
    \infer[\RuleINSTIMPL]{%
      \CalledInstanceImplMethod{i}{m}{\lambda}
    }{
      \LambdaCGE{*}{m}{\lambda} \ruleAnd{}
      \neg\IsStatic{m} \ruleAnd{}
      \LambdaObject{\lambda}{*}{*}{i}
    }
  \end{tabular}\\[1em]

  \begin{tabular}{c}
    \infer[\RuleSHIFT{1}]{%
      \ParamsReceiverShiftRight{\lambda}{m}{0}{0}
    }{
      \LambdaObject{\lambda}{m}{*}{*} \ruleAnd{}
      \IsStatic{m}
    }
  \end{tabular}\\[1em]

  \begin{tabular}{cc}
    \infer[\RuleSHIFT{2}]{%
      \ParamsReceiverShiftRight{\lambda}{m}{0}{1}
    }{
      \CalledInstanceImplMethod{i}{m}{\lambda} \ruleAnd{}
      \Arity{i} = 0
    } &
    \infer[\RuleSHIFT{3}]{%
      \ParamsReceiverShiftRight{\lambda}{m}{1}{0}
    }{
      \CalledInstanceImplMethod{i}{m}{\lambda} \ruleAnd{}
      \Arity{i} > 0
    }
  \end{tabular}\\[1em]

  \begin{tabular}{c}
    \infer[\RuleLARGS]{%
      \VPT{v}{\val}
    }{
      \LambdaCGE{i}{m}{\lambda} \ruleAnd{}
      \ParamsReceiverShiftRight{\lambda}{m}{k}{n} \ruleAnd{}
      \LambdaObject{\lambda}{m}{*}{i} \hspace{1.5cm} \\
      \Act{i}{n'} = v' \ruleAnd{} 
      \Frm{m}{n''} = v \ruleAnd{} 
      n'' = \Arity{i}-(k+n)+n' \ruleAnd{}
      \VPT{v'}{\val}
    }
  \end{tabular}\\[1em]

  \begin{tabular}{c}
    \infer[\RuleCAPTARGS]{%
      \VPT{\Frm{m}{n-k}}{\val}
    }{
      \LambdaCGE{*}{m}{\lambda} \ruleAnd{}
      \ParamsReceiverShiftRight{\lambda}{m}{k}{*} \ruleAnd{}
      \LambdaObject{\lambda}{m}{*}{i} \ruleAnd{}
      \LambdaCaptured{i}{n}{\val} \ruleAnd{}
      k + n \le \Arity{i}
    }
  \end{tabular}\\[1em]

  \begin{tabular}{c}
    \infer[\RuleLAMBDATHIS]{%
      \VPT{\ThisVar{m}}{\val}
    }{
      \ParamsReceiverShiftRight{\lambda}{m}{1}{0} \ruleAnd{}
      \CalledInstanceImplMethod{i}{m}{\lambda} \ruleAnd{}
      \LambdaCaptured{i}{0}{\val} \ruleAnd{}
    }
  \end{tabular}\\[1em]

  \begin{tabular}{c}
    \infer[\RuleMREFTHIS]{%
      \VPT{\ThisVar{m}}{\val}
    }{
      \LambdaCGE{i}{m}{\lambda} \ruleAnd{} \ruleAnd{}
      \ParamsReceiverShiftRight{\lambda}{m}{0}{1} \ruleAnd{}
      \VPT{\Act{i}{0}}{\val}
    }
  \end{tabular}\\[1em]

  \begin{tabular}{c}
    \infer[\RuleCCALL]{%
      \VPT{v}{\val} \ruleAnd{}
      \VPT{\ThisVar{m}}{\val}
    }{
      \LambdaCGE{i}{m}{\lambda} \ruleAnd{}
      \IsConstructor{m} \ruleAnd{}
      m \in t \ruleAnd{} \ruleAnd{}
      \AssignRet{i} = v \ruleAnd{}
      \val = \mockC{t}{i}
    }
  \end{tabular}\\[1em]

  \end{centering}
\caption{Rules for handling method references and lambdas.}
\label{fig:lambdas}
\end{figure}

\section{Evaluation}
\label{sec:evaluation}

We evaluate our analysis on two test suites: a microbenchmark suite of
our own (Section~\ref{sec:microbenchmark-suite}) and the test suite of
\emph{Sui et al.}~\cite{Sui18}
(Section~\ref{sec:dynamic-benchmark-suite}).

Our analysis is implemented in the declarative static analysis
framework Doop~\cite{oopsla/BravenboerS09}. All analyses are run on a
64-bit machine with an Intel Xeon CPU E5-2667 v2 3.30GHz with 256 GB
of RAM. We use the Souffl\'e compiler (v.1.4.0), which compiles
Datalog specifications into binaries via C++ and run the resulting
binaries in parallel mode using four jobs. Doop uses the Java 8
platform as implemented in Oracle JDK v1.8.0\_121. All running times and
precision numbers are for Doop's default context-insensitive analysis.
(Context sensitivity adds no precision to the high-level metrics shown.)
For benchmarks of generalized \invokedynamic{} features (i.e., not lambdas
and method references), we enable reflection support in Doop.


\subsection{Microbenchmark Suite}
\label{sec:microbenchmark-suite}

To evaluate our technique, we have built our own suite of
microbenchmarks. 
These benchmarks capture a large number of idioms found in realistic uses of method
references (Section~\ref{sec:method-references-benchmark}), lambdas
(Section~\ref{sec:lambdas-benchmark}), and method handles combined
with \invokedynamic{} (Section~\ref{sec:invokedynamic-benchmark}), including
most of the patterns shown in the examples of Section~\ref{sec:examples}. (Other
patterns are captured in the Sui et al. suite, discussed later.)
The suite is freely available.

Analysis times for the three component benchmarks
are shown in Figure~\ref{fig:microbenchmark}. As can be seen, enabling
reflection analysis, for fully general handling of \invokedynamic{},
incurs higher cost.

\begin{figure}
  \begin{center}
\begin{tabular}{|l|r|}
  \hline
  \textbf{Benchmark}                           & \textbf{Time (sec)} \\
  \hline
  \textbf{Method References}                   & 27                  \\
  \textbf{Lambdas}                             & 23                  \\
  \textbf{Method Handles and \invokedynamic{}} & 378                 \\
  \hline
\end{tabular}
  \end{center}
\caption{Microbenchmark times.}
\label{fig:microbenchmark}
\end{figure}

Our static analysis fully models all behavior in the microbenchmark
suite. Although the suite was developed in tandem with the analysis,
it still provides partial validation of analysis completeness, given the
effort to encode many variations of operations, as detailed next.

\subsubsection{Microbenchmark: Method References}
\label{sec:method-references-benchmark}

This benchmark includes Oracle's tutorial code
MethodReferencesTest~\cite{methodreferencestutorial}. We capture the
behavior of all four kinds of methods references (found in the
tutorial table): to static methods, to instance methods of a
particular object, to instance methods of an arbitrary object of a
particular type, and to constructors.

The microbenchmark also contains code that showcases 
the following features:
\begin{enumerate}
\item Construction of functional objects directly from method references.
\item Use of functional objects together with Java 8 stream API methods.
\item Auto-boxing conversions.
\end{enumerate}

\subsubsection{Microbenchmark: Lambdas}
\label{sec:lambdas-benchmark}

This benchmark shows the handling of the following features:

\begin{enumerate}
\item Creating lambdas with arrow notation. This includes nested lambdas.
\item Creating lambdas that can access values of the outside
  environment (forming closures).
\end{enumerate}

\subsubsection{Microbenchmark: Method Handles and \invokedynamic{}}
\label{sec:invokedynamic-benchmark}

Java currently does not support the direct representation of
\invokedynamic{} in source code, although such a feature is 
considered for inclusion in future versions of the
language~\cite{jep303}. For this reason, this benchmark uses the ASM
bytecode manipulation library\footnote{\url{https://asm.ow2.io/}} to
dynamically generate and load a class with \invokedynamic{}
invocations.

The benchmark captures the following patterns:
\begin{enumerate}
\item Lookup of a \texttt{MethodHandles.Lookup} object via
  \texttt{MethodHandles.lookup()}.
\item Construction of method type values via
  \texttt{MethodType.methodType()} methods.
\item Look-up of virtual and static methods via
  \texttt{MethodHandles.Lookup.findVirtual()} and
  \texttt{MethodHandles.Lookup.findStatic()}.
\item Calling method handles with \texttt{MethodHandle.invokeExact()}.
\item Passing a receiver for non-static methods (thus handling places
  where the signature of the target method differs from the signature
  of the \texttt{MethodHandle.invokeExact()} signature found in the
  bytecode).
\item Bootstrapping calls to another class in a manner similar to the
  motivating example in Section~\ref{sec:example-late-linking}.
\end{enumerate}

\subsection{Sui et al. Test Suite}
\label{sec:dynamic-benchmark-suite}

We also evaluate our technique using the
dynamic features test suite of \emph{Sui et al.}~\cite{Sui18}. This is
a test suite that examines the soundness of call-graph construction
and is written to specifically test the static analysis of features
such as lambdas and \invokedynamic{}, by authors with extensive
experience in systematic Java benchmarking efforts
(e.g., XCorpus~\cite{Dietrich17b}).


The benchmark suite contains three benchmarks for lambdas, plus a
benchmark for \invokedynamic{} in general (Dynamo).  Dynamo is a
realistic software artifact~\cite{Jezek16} that has been configured in
the benchmark suite to specifically evaluate the analysis of
\invokedynamic{}. The Dynamo library exercises all features of dynamic
invocation sites (static vs. non-static, constructors, signature
adaptation, interaction with plain Java reflection). It injects
\invokedynamic{} calls in unsuspecting code to address cross-component
linking errors. Thus, if these \invokedynamic{} sites are not
analyzed, then the static analysis cannot find calls from code to a
library.

The Dynamo test program in the suite contains two \invokedynamic{} sites:
\begin{enumerate}
\item A site that looks up a constructor method and creates an
  object. Since it is a constructor method handle, the analysis also
  recognizes that an object must also be allocated for this
  invocation.
\item A site that looks up an instance method and calls it. The
  original signature of the method accepts an object and is adapted to
  also accept the receiver.
\end{enumerate}
\noindent In both cases, Dynamo retrieves the method via reflection
and then proceeds to ``unreflect'' it. The test program does not test
lookup of static methods.

For every benchmark, the following ground truth is provided: one or more methods
are expected to be found reachable, while one or more different
methods are expected to be found unreachable. The results of applying
our analysis to these benchmarks are shown in
Figure~\ref{fig:dynbenchmark}.

\begin{figure}
  \begin{center}
\begin{tabular}{|l|ll|ll|r|}
  \hline
  \textbf{Benchmark}      & \textbf{Reachable} &          & \textbf{Unreachable} &          & \textbf{Time (sec)} \\
  \hline
                          & expected           & analysis & expected             & analysis &                     \\
  \hline
  \textbf{LambdaConsumer} & 1                  & \ok{}    & 1                    & \ok{}    & 21                  \\
  \textbf{LambdaFunction} & 1                  & \ok{}    & 2                    & \ok{}    & 21                  \\
  \textbf{LambdaSupplier} & 1                  & \ok{}    & 1                    & \ok{}    & 22                  \\
  \textbf{Dynamo}         & 1                  & \ok{}    & 1                    & \no{}    & 242                 \\
  \hline
\end{tabular}
  \end{center}
\caption{Dynamic benchmark results.}
\label{fig:dynbenchmark}
\end{figure}

Notably:
\begin{itemize}
\item All lambda benchmarks are analyzed precisely: the expected
  methods are found reachable or unreachable.
\item For Dynamo, our analysis over-approximates
  reachability. Dynamo uses
  \invokedynamic{} as a layer between components to ensure binary
  compatibility with evolving code.
  As seen in Figure~\ref{fig:dynbenchmark}, our analysis
  over-approximates reachability: it discovers the expected method as
  reachable but also discovers the expected unreachable method. This
  problem is not fundamental to the technique that we present, but is
  caused by the lack of flow sensitivity in the underlying points-to
  analysis, provided by the Doop framework. Dynamo code creates method
  handles by gathering reflectively all members of classes and then
  selectively filtering out the ones that do not match; Doop's flow
  insensitivity causes it to ignore this filtering. Coupled with flow
  sensitivity, our technique should be able to ignore the expected
  unreachable method.
\item The efficiency of a lambda-specialized analysis vs. a
  general-purpose \invokedynamic{} analysis that requires
  reflection support is again demonstrated in the running times.
\end{itemize}

\section{Related work}
\label{sec:related}


\subparagraph*{Static Analysis of Java Lambdas and Dynamic Calls.}
Some recent work has
attempted to treat lambdas and their static analysis, mostly in isolation,
as another high-level feature for practical tools. Cifuentes~\emph{et
  al.}~\cite{Cifuentes15} perform a pattern-based vulnerability
analysis (i.e., not a full low-level analysis of value flow) and recognize
code patterns containing lambdas. There
has also been work on dynamic analyses that understand Java-style
lambdas~\cite{Erdweg14}.

Reflection and programmable dynamic calls are subtle features that should
be formalized in order to be addressed. However, the bibliography is
lacking: we only know of the work of Landman~\emph{et
  al.}~\cite{Landman17}, who give a syntax of the DSL behind the
standard Java Reflection API. They do not treat its semantics, as they
did not need to (their work was on mining big codebases for the
existence of specific patterns).

To the best of our knowledge, no formal semantic model of
\invokedynamic{} and its API exists.
Other Java APIs that cannot be easily analyzed
statically have also been candidates for static semantic
modeling. Smaragdakis~\emph{et al.} model the reflection
API~\cite{aplas/Smaragdakis15} and Fourtounis~\emph{et al.} model
dynamic proxies~\cite{Fourtounis18}. Our approach differs in two
aspects: (a)~we do not necessarily incur performance overheads (our
handling of functional objects does not require expensive reflection
support) and (b)~we model the lower-level \texttt{java.lang.invoke}
API, which requires handling of JVM features such as signature
polymorphism, caller sensitivity, and reasoning about code running at
class-loading time.




The IBM WALA static analysis framework~\cite{www:wala}
has limited support for \invokedynamic{}, specifically for
call-graph edges over lambdas by generating synthetic classes.\footnote{\url{https://groups.google.com/forum/\#!topic/wala-sourceforge-net/omsGtp\_ow7I},\\ \url{https://github.com/wala/WALA/blob/f2b1e9fec0627e221427404cb7ba194c4a89cd9e/com.ibm.wala.core/src/com/ibm/wala/ipa/summaries/LambdaSummaryClass.java\#L42}}
WALA also lacks full support for constructor method references~\cite{Reif18}.

\subparagraph*{Transforming Away \invokedynamic{}.}
Lambdas are not easy to work with; Soot, a popular Java manipulation
and analysis framework, even considers statically transforming them
away~\cite{Soot226}, since
\invokedynamic{} has been too difficult to analyze:
``Soot does not fully support dynamic invokes ... could not find an
easy workaround and instead decided that it would be best to change
Schaapi such that dynamic invokes (and thus lambdas) are ignored
completely.''\footnote{\url{https://github.com/cafejojo/schaapi/pull/295}}

Along the same lines, but more completely, the OPAL bytecode
rectifier\footnote{\url{http://www.opal-project.de/DeveloperTools.html}}
removes instances of \invokedynamic{} as used in Java lambdas. This is
a general alternative static treatment of lambdas, but not of other
instances of \invokedynamic{}.  Similar removal of stylized uses of
\invokedynamic{}, without handling the general case, are performed by
RetroLambda\footnote{\url{https://github.com/luontola/retrolambda}}
and Google's
D8.\footnote{\url{https://jakewharton.com/androids-java-8-support/}}
These tools cannot, e.g., make the Dynamo benchmark analyzable by
analyses that do not understand \invokedynamic{}.


\subparagraph*{Other Platforms.}

Apart from the popular OpenJDK and its VM, used on servers or
desktops, the other mainstream Java platform is Android. The
implementation of \invokedynamic{} on Android posed some complications
because dynamic code generation is restricted on Android due to
resource constraints~\cite{Pilliet15,Roussel14}. 
\invokedynamic{} was prototyped
for Android~\cite{Roussel14} and, eventually,
became officially supported when the latest ``Android N'' switched to
Java 8. Our work is, thus, applicable to Android as well. Android
is a platform that commands special attention due to its popularity.
\invokedynamic{}
enables new optimizations and analyses~\cite{Xu15,Xu16,Xu17}. However, the
instruction is also a security threat, since it is so powerful that it
can, for example, hide method calls and make malware undetectable (as
demonstrated by the DexProtector tool~\cite{dexprotector} or the
survey of Gorenc and Spelman~\cite{everydays}) and provides less
security by-design compared to classic
reflection~\cite{security12}.

The .NET platform also has functionality similar to method handles and
anonymous classes, called ``dynamic methods''~\cite{dynamic_method}.
We, thus, expect that our approach
can be ported to other runtimes and to their implementation of
dynamic features.




\section{Conclusion}
\label{sec:conclusion}

We presented a static analysis modeling of programmable dynamic
linking in Java, i.e., the \invokedynamic{} instruction and
accompanying framework.  The approach addresses the most fundamental
level of the language feature, fully modeling method handles, while at
the same time it maintains high efficiency and completeness for common
uses of \invokedynamic{} in Java lambdas. This is the first
thorough handling of the \invokedynamic{} feature, which had
so far resisted static analysis.

\bibliography{references,bib/ptranalysis,bib/proceedings,bib/tools}

\end{document}